\documentclass[revtex4]{emulateapj}

\usepackage{graphicx}

\begin{document}

\shorttitle{Primary CRs} \shortauthors{Dogiel et al.}

\title{Formation of the cosmic-ray halo: Galactic spectrum of primary cosmic rays}
\author{V.~A.~Dogiel$^1$, A.~V.~Ivlev$^2$, D.~O.~Chernyshov$^{1}$, C.-M.~Ko$^3$ }
\affil{$^1$I. E. Tamm Theoretical Physics Division of P. N. Lebedev Institute of Physics, Leninskii pr. 53, 119991 Moscow,
Russia} \affil{$^2$Max-Planck-Institut f\"{u}r Extraterrestrische Physik, Giessenbachstr. 1, D-85748 Garching, Germany}
\affil{$^3$Institute of Astronomy, National Central University, Zhongli Dist., Taoyuan City, Taiwan (R.O.C.)}

\altaffiltext{*}{e-mail: ivlev@mpe.mpg.de}

\begin{abstract}
A self-consistent model of a one-dimensional cosmic-ray (CR) halo around the Galactic disk is formulated with the
restriction to a minimum number of free parameters. It is demonstrated that the turbulent cascade of MHD waves does not
necessarily play an essential role in the halo formation. Instead, an increase of the Alfv\'{e}n velocity with distance to
the disk leads to an efficient generic mechanism of the turbulent redshift, enhancing CR scattering by the self-generated
MHD waves. As a result, the calculated size of the CR halo at lower energies is determined by the halo sheath, an
energy-dependent region around the disk beyond which the CR escape becomes purely advective. At sufficiently high energies,
the halo size is set by the characteristic thickness of the ionized gas distribution. The calculated Galactic spectrum of
protons shows a remarkable agreement with observations, reproducing the position of spectral break at $\approx0.6$~TeV and
the spectral shape up to $\sim10$~TeV.
\end{abstract}



\maketitle
\section{Introduction}\label{1}

There have been a great deal of attempts to formulate adequate models of cosmic rays (CRs) in the Galaxy, starting from
pioneer papers by Ginzburg and Syrovatskii, as summarized in their monograph \citep[][]{syr64}. The idea of an extended CR
halo around the Galactic disk was proposed by \citet{vl53}. Phenomenological models of CR propagation by diffusion in the
halo, with free escape at the halo edges, had been developed by \citet{syr59,bul72a, bul76,ptus72,ptus74,
Ginzburg1973,bul74, Hillas1975}, and the kinetic theory was further advanced by \citet{Jokipii1966,Jokipii1967,
Gleeson1974,Webb1974, Owens1977,Lerche1982} -- of course, the list above cannot be comprehensive.

Several generations of gamma-ray telescopes -- from COS-B and EGRET \citep{Harding1985,Mayer1987, Strong1988,Zhang1994,
Strong1996} to Fermi-LAT \citep{Acero2016a,Acero2016b, recc16,Yang2016} -- have later provided important constraints on the
CR distribution in the Galactic disk. First observations indicated rather weak inhomogeneities of CRs in the disk,
suggesting a global CR mixture. Sophisticated numerical codes -- such as GALPROP \citep[see, e.g.,][]{mosk98,vlad11,joh18}
as well as DRAGON \citep[][]{Gaggero2013}, PICARD \citep[][]{Kissmann2014}, and CRPropa \citep[][]{AlvesBatista2016,
Merten2017} -- have been developed for interpretations of CR data and nonthermal emission in the Galaxy.

Phenomenological models of the CR halo provide a detailed picture of essential processes occurring in the Galaxy, and also
advance our understanding of the global problem of CR origin \citep[see, e.g.,][]{Ginzburg1972,Ginzburg1976,
Cesarsky1980,Dogiel1989, Aloisio2013,Amato2018,tjus20}. However, several critical parameters remain beyond the scope of the
models. One of such parameters of fundamental importance is the size of the halo, i.e., the height $z_{\rm H}$ where CRs
escape from the Galaxy. The energy spectrum of CRs in the Galactic disk is inherently related to the halo size and its
dependence on the CR energy.

One can identify three distinct classes of models describing the CR halo:
\begin{itemize}
\item {\it Static halo.} These models assume that CRs are confined by scattering on magnetic fluctuations, which leads
    to CR diffusion within the halo and, irrespective of their kinetic energy $E$, allows free escape at a given halo
    edge \citep[see][]{syr64, ber90,blasi19}. The spatial distribution of CR sources plays a minor role in such models.
    The main phenomenological parameters are the spatial diffusion coefficient of CRs $D(E)$ and the halo size $z_{\rm
    H}$; recently, \citet{bosch17,bosch20} estimated $D\sim10^{28}(E/1~{\rm GeV})^{0.3}$~cm$^2$~s$^{-1}$ and $z_{\rm
    H}\approx4$~kpc as input parameters for the GALPROP code.
\item {\it Advection halo.} Static models can be extended by adding CR advection with velocity $v_{\rm adv}$ to the
    diffusion flux. This takes into account that escaping CRs generate MHD waves which, in turn, provide their
    scattering \citep{Jones1979, Lerche1982,bloemen:93, breit02,Blasi2012}. The escape boundary is then located at a
    height of $\sim D(E)/v_{\rm ad}\lesssim3$~kpc, which is smaller than $z_{\rm H}$ in static models. For CRs which are
    frozen in MHD fluctuations (propagating with the Alfv\'{e}n velocity $v_{\rm A}$), the flux velocity of escaping CRs
    is about $v_{\rm adv}\approx v_{\rm A}$ if their transport is dominated by advection and excited waves primarily
    propagate outward \citep[see, e.g.,][]{wentzel:74,Jones1993,comm19}.
\item {\it Nonuniform halo.} These (partially) self-consistent models can predict the halo size, which generally depends
    on the CR energy. Such a model was first proposed by \citet{dogiel:93}, who assumed the diffusion coefficient to be
    a given function of energy. A nonlinear self-consistent extension of this model, including the kinetic equation for
    CR-excited MHD waves, was reported in \citet{dogiel:94} where $z_{\rm H}(E)$ was derived analytically. Recently, a
    nonlinear model of the CR halo was developed by \citet{evoli18}, whose numerical analysis suggests that the halo
    arises due to MHD turbulence cascading to larger wavenumbers. This reduces CR scattering with height, and eventually
    sets free CR escape at a certain energy-independent $z_{\rm H}$.
\end{itemize}

One should also mention MHD models of the halo, proposed by \citet{breit:91,breit:93} and further developed by, e.g.,
\citet{recc16,buck20,holg19}. These models take into account back-reaction of escaping CRs on the gas, which can lead to the
onset of hydrodynamic outflows from the disk. In the present paper we focus on generic mechanisms of CR halo formation, and
therefore assume the ambient gas to be at rest.

The principal goal of this paper is to formulate a self-consistent one-dimensional model of the CR halo, which is
restricted to a minimum number of free parameters. We demonstrate that the turbulent cascade of MHD waves, which some models
consider to play an essential role in forming the halo, may in fact be insignificant. Instead, an increase of the Alfv\'{e}n
velocity with height leads to an efficient mechanism of the turbulent {\it redshift}, enhancing CR scattering by the
self-generated MHD waves. The ultimate halo edge in our model is therefore set by a transition to Alfv\'{e}nic advection of
escaping CRs.

The paper is organized as follows. In Section~\ref{2} we summarize available observational constraints relevant for our
problem, determine the main input parameters of the model, and formulate the governing equations describing coupled spectra
of CRs and MHD waves outside the Galactic disk. In Section~\ref{mech} we present an approximate solution of the governing
equations and discuss the mechanism resulting in the formation of the CR halo. We introduce a halo sheath -- a region around
the disk beyond which the CR transport is ultimately dominated by Alfv\'{e}nic advection, and analyze the relative
importance of the excited turbulence in shaping the CR spectrum in the disk. In Section~\ref{nsol} the governing equations
are solved numerically for two characteristic models of ionized gas distribution in the halo, which allows us to compare our
results with observational data on the proton spectra in the disk. Our results show a remarkable agreement with observations
at energies $\gtrsim50$~GeV (where the solar modulation and re-acceleration are completely negligible) and up to
$\sim10$~TeV (beyond which the data quality is worsening). Finally, in Section~\ref{discuss} we analyze limitations of our
approach, compare our model of a nonuniform halo with other models proposed previously, and summarize the obtained results.

\section{Model parameters and governing equations}\label{2}

Parameters of our model are primarily constrained by measurements of CR properties inside the Galaxy \citep[see,
e.g.,][]{tjus20}.
\begin{itemize}
\item Theoretical models of CR acceleration predict that the source energy spectrum of CRs injected by supernova
    remnants (SNRs) obeys a power law $\propto E^{-\gamma}$, with the spectral index of $\gamma=2$ for strong shocks
    \citep[][]{krym77,bell78}. Recent gamma-ray observations suggest $\gamma=2.3$ for CRs injected by young SNRs, such
    as Tycho, CasA, etc. \citep[][]{tjus20}.
\item The total CR luminosity in the Galaxy, $L_{\rm CR}$, determines the magnitude of CR sources. Different methods
    applied to available observational data suggest $L_{\rm CR}\sim 10^{40}$~erg~s$^{-1}$ \citep{strong10,mura19,
    tjus20}. The differential flux of relativistic CR protons from Galactic disk can then be evaluated as
    \begin{eqnarray}
    S(E)\approx\frac{L_{\rm CR}}{2\pi R_{\rm disk}^2(m_pc^2)^2}\left(\frac{m_pc^2}{E}\right)^{\gamma}\nonumber\hspace{1.4cm}\\
    \equiv S_*\left(\frac{m_pc}{p}\right)^{\gamma},\label{source}
    \end{eqnarray}
    where $R_{\rm disk}\approx15$~kpc the radius of Galactic disk, $S_*\approx
    5\times10^{-4}$~cm$^{-2}$~s$^{-1}$~GeV$^{-1}$ is the flux scale, and $p$ is the momentum of a proton of mass $m_p$.
\item The spatial diffusion coefficient of relativistic CRs, $D_0$, in the Galactic plane $z=0$ is estimated from the
    abundance of CR nuclei. By measuring the boron-to-carbon abundance ratio B/C versus momentum yields $({\rm
    B/C})\propto p^{-\delta}$ with $\delta=0.33$ \citep{agu16}. For relativistic particles, this gives the
    dependence $D_0(p)\propto p^{\delta}$. The latter suggests a power-law spectrum of MHD fluctuations in the disk,
    $W_0(k)\propto k^{-\beta}$, where $\beta=2-\delta$ and $k$ is the wavenumber. Thus, the diffusion coefficient
    of relativistic protons in the Galactic plane versus their momentum can be presented as
\begin{equation}\label{D_disk}
    D_0(p)=D_*\left(\frac{p}{m_pc}\right)^{2-\beta},
\end{equation}
    where $D_*\sim 10^{28}$~cm$^2$~s$^{-1}$ \citep[e.g.,][]{evoli18}.
\item The gas distribution in the halo was estimated in \citet{gaen08,sun10,mill16}, showing that the halo at
    $z\sim1$~kpc is filled by warm ionized gas with the average density of $\sim0.01$~cm$^{-3}$. The gas scale height of
    $\approx2$~kpc was estimated from Low-Frequency Array (LOFAR) \citep[see][and also \citealt{avil07,faber18}]{sob19}.
    For numerical calculations below (Section~\ref{nsol}), we adopt two different models of gas distribution. One model,
    denoted by CL02, assumes a superposition of ``thin'' and ``thick'' disks \citep[see][]{cord03}. In this case, the
    number density of ionized gas (in cm$^{-3}$) is given by a sum of the respective contributions,
\begin{eqnarray}
    n(z) = 0.1\cosh^{-2}\left(\frac{z}{0.14~{\rm kpc}}\right)\nonumber\hspace{1.4cm}\\
    +0.035\cosh^{-2}\left(\frac{z}{0.95~{\rm kpc}}\right),\label{n12}
\end{eqnarray}
    Equation~(\ref{n12}) is illustrated in Figure~\ref{fig1} by the solid line. The second model, MB13, is a power-law
    profile \citep[see][]{mill16},
\begin{equation}\label{beta}
    n(z) = 0.46\left[1 + \left(\frac{z}{0.26~\mbox{kpc}}\right)^2\right]^{-1.1},
\end{equation}
    which is depicted in Figure~\ref{fig1} by the dotted line.
\begin{figure}[!h]
\begin{center}
\resizebox{\hsize}{!}{\includegraphics{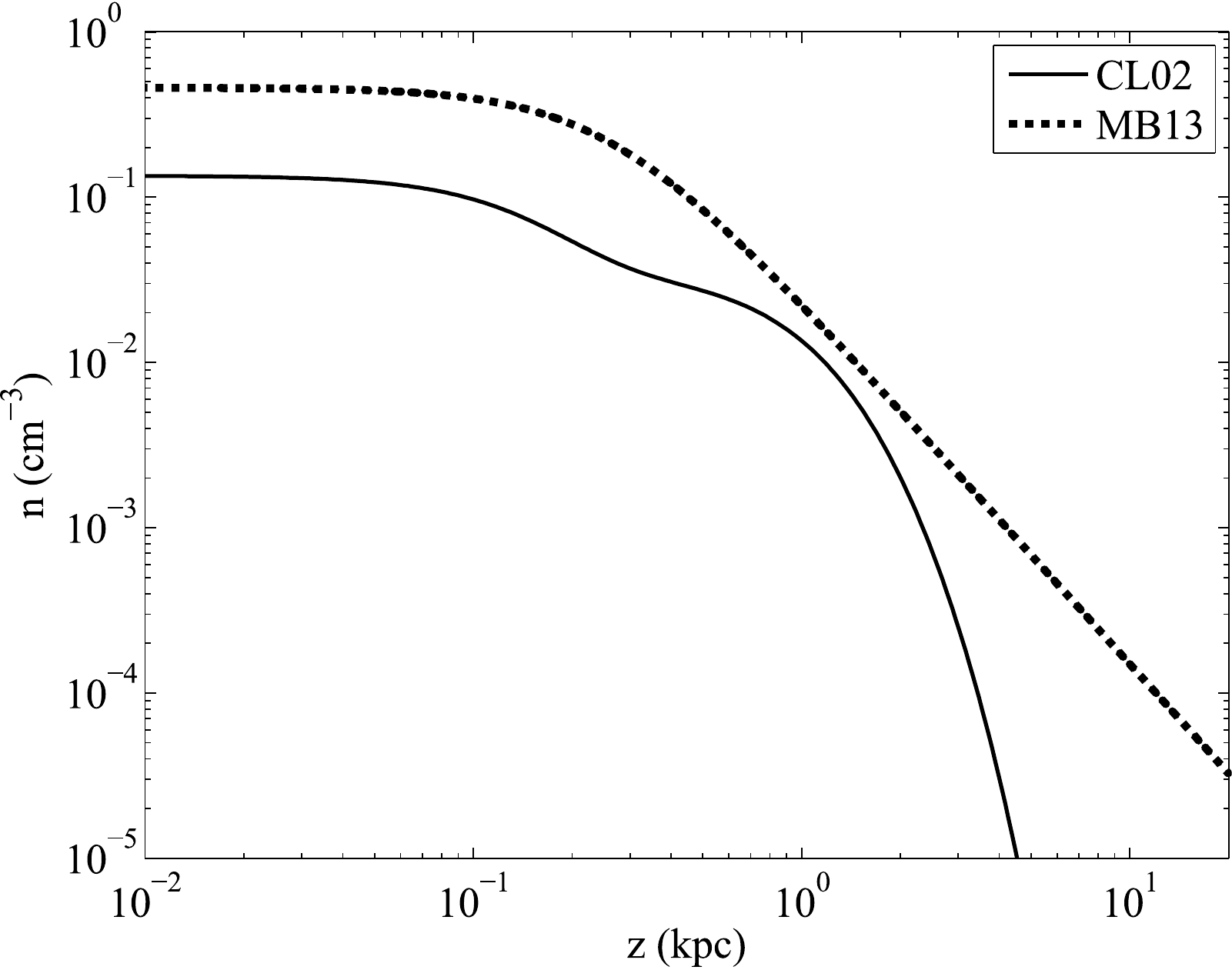}}
\caption{Characteristic model distributions of ionized gas in the halo. The solid line shows the exponential profile CL02,
Equation~(\ref{n12}); the dotted line represents the power-law profile MB13, Equation~(\ref{beta}).} \label{fig1}
\end{center}
\end{figure}
\item The magnetic field above the Galactic disk is, generally, nonuniform. In the halo, it can be represented by
    diverging field lines with the characteristic scale height of about 10~kpc \citep[][]{breit:91,breit:93,
    zir:96,dorf12,dorf19}.
    Equations~(\ref{n12}) and (\ref{beta}) show that the gas scale height in the halo is substantially smaller, and
    thus the height dependence of the Alfv\'{e}n velocity,
\begin{equation}
    v_{\rm A}(z)=\frac{B(z)}{\sqrt{4\pi mn(z)}}\,,
\end{equation}
    is primarily determined by $n(z)$ (here $m\approx m_p$ is the ion mass). Below we show that the gas scale height
    essentially sets the maximum size of the CR halo in our model. Therefore, in the analysis below we can assume the
    magnetic field of a constant strength $B$ of about several $\mu$G.
\end{itemize}

Our analysis rests on two governing equations describing CRs in the halo, {\it outside the disk}. The steady-state transport
equation for the CR spectral density, $N(p,z)$, has the standard form \citep[e.g.,][]{ber90,skil1},
\begin{equation}
\frac{\partial}{\partial z}\left(v_{\rm A}N-D\frac{d N}{\partial z}\right)-\frac{1}{3}\frac{d v_{\rm A}}{dz}\,
\frac{\partial}{\partial p}\left(pN\right)=0\,,
\label{tdif}
\end{equation}
where $v_{\rm A}(z)$ is determined by $n(z)$. The spectral density in momentum space $N(p)$, related to that in energy space
via $N(p)=v(E)N(E)$, is normalized such that $\int N(p)\,dp$ is the total number density of CRs. The diffusion coefficient
$D$ of a CR particle with the physical velocity $v$ is determined using the approximation by \citet{skil3},
\begin{equation}
D(p,z)\approx\frac{vB^2}{6\pi^2k^2W}\,,
\label{ddif}
\end{equation}
where $W(k,z)$ is the energy density of MHD fluctuations at the resonant wavenumber $k$. The latter is related to the CR
momentum $p$ via the resonance condition,
\begin{equation}\label{resonant}
kp\approx m_p\Omega_*\,,
\end{equation}
where $\Omega_*=eB/m_pc$ is the gyrofrequency scale of CR protons.

In Equation~(\ref{tdif}) we omitted terms describing (i) the second-order Fermi acceleration of CRs by the disk turbulence,
and (ii) energy losses due to ionization/Coulomb collisions. Both terms may affect the CR spectrum at energies below
$\sim$~GeV and therefore are not essential for our problem. Furthermore, we note that re-acceleration is characterized by
the diffusion coefficient in momentum space $D_{pp}$, which is related to the spatial diffusion coefficient via
$D_{pp}D\approx\frac19p^2v_{\rm A}^2$ \citep[for isotropic turbulence, see, e.g.,][]{Thornbury2014} and hence does not
introduce new parameters into the problem. Similarly, the energy losses (dominated by Coulomb collisions in a fully ionized
halo) are completely characterized by the density of ionized gas.

The equation for MHD turbulence in the halo is generally derived from Liouville's theorem for the phase density of MHD
disturbances $\mathcal{N}({\bf k},{\bf r})$ in a nonuniform medium \citep[see][]{dogiel:94}. Using the Hamiltonian equations
of motion, $\dot{\bf r}=\partial\omega/\partial {\bf k}$ and $\dot{\bf k}=-\partial\omega/\partial {\bf r}$, we have
\begin{equation}\label{Liu1}
\frac{\partial {\mathcal N}}{\partial t}+\frac{\partial {\mathcal N}}{\partial {\bf r}}\cdot
\frac{\partial \omega}{\partial {\bf k}} -\frac{\partial {\mathcal N}}{\partial {\bf k}}\cdot
\frac{\partial \omega}{\partial {\bf r}}={\mathcal S}-{\mathcal L}\,,
\end{equation}
where ${\mathcal S}$ and ${\mathcal L}$ are the terms representing relevant sources and losses, and $\omega({\bf k},{\bf
r})$ is the wave frequency. The phase density is related to the MHD energy density $W({\bf k},{\bf r})$ via
$\mathcal{N}=W/\omega$. We substitute this in Equation~(\ref{Liu1}) and take into account that we consider a one-dimensional
turbulence for transverse MHD waves propagating along the magnetic field (Alfv\'{e}n and fast magnetosonic modes,
efficiently contributing to the CR scattering). Using the dispersion relation $\omega(k,z)=v_{\rm A}(z)k$, after simple
manipulation we obtain the following equation for stationary energy density of waves in the halo:
\begin{equation}
\frac{\partial}{\partial z}(v_{\rm A}W)-\frac{d v_{\rm A}}{d z}\,\frac{\partial}{\partial k}(kW)=2\Gamma W\,,
\label{vaw}
\end{equation}
where the source term on the r.h.s.\ is determined by the {\it amplitude rate} of resonant wave excitation, $\Gamma$, while
explicit loss terms are omitted (see Appendix~\ref{notes}). In the approximation by \citet{skil3}, the excitation rate is
proportional to the CR diffusion flux,
\begin{equation}
\Gamma(k,z)\approx-\frac{\pi^2e^2v_{\rm A}p}{m_pc^2\Omega_*}\,D\frac{\partial N}{\partial z}\,.
\label{gmm}
\end{equation}
Equation~(\ref{vaw}) is written for waves propagating upward, in the direction of CR flux. Downward-propagating waves are
damped at the same rate.

We point out that the second term on the l.h.s.\ of Equation~(\ref{vaw}) is usually absent in wave equations describing
CR-induced MHD turbulence. Furthermore, the derived equation does not contain explicit damping terms as well as nonlinear
terms representing interactions between waves. In Appendix~\ref{notes} we discuss these points in detail.

\section{Mechanism of the CR halo formation}
\label{mech}

In this Section we rewrite the governing equations in a dimensionless form, derive their asymptotic solution, and obtain an
approximate analytical solution for the CR spectrum in the Galactic disk.

\subsection{Dimensionless equations}

For simplicity, we assume an exponential gas density profile,
\begin{equation}\label{n_exp}
n(z)=n_0e^{-z/z_n}\,,
\end{equation}
so that the Alfv\'{e}n velocity is $v_{\rm A}(z)=v_{\rm A}^0e^{\tilde z}$ with the normalized vertical coordinate (height)
$\tilde z=z/2z_n$. Dimensionless CR spectral density in {\it momentum} space, $\tilde N(p)=v_{\rm A}^0N(p)/vS_{*}$, is
normalized by the characteristic density of CR sources (\ref{source}) in the advection regime, $S_{*}/v_{\rm A}^0$. The
resonance condition (\ref{resonant}) for dimensionless momentum and wavenumber, $\tilde p=p/m_pc$ and $\tilde
k=kc/\Omega_*$, respectively, is reduced to $\tilde p\tilde k=1$, so that
\begin{equation}\label{res}
\ln \tilde p=-\ln \tilde k \equiv q\,.
\end{equation}
We assume relativistic protons (i.e., $q\gtrsim0$) and set $v\approx c$ in the expression for $D(p,z)$. This allows us to
write governing equations (\ref{tdif}) and (\ref{vaw}) for the CR density and energy density of resonant MHD turbulence in
the following dimensionless form:
\begin{eqnarray}
\frac{\partial}{\partial \tilde z}\left(-\tilde D\frac{e^{\tilde z+2q}}{\tilde U}\,\frac{\partial \tilde N}{\partial \tilde z}
+e^{\tilde z+q}\tilde N\right)=\frac13\frac{\partial}{\partial q}\left(e^{\tilde z+q}\tilde N\right), \hspace{.5cm}\label{N}\\
\frac{\partial \tilde U}{\partial \tilde z}+\frac{\partial \tilde U}{\partial q}
=-e^{\tilde z+2q}\frac{\partial \tilde N}{\partial \tilde z}\,.\hspace{.5cm}\label{U}
\end{eqnarray}
Here we introduced an auxiliary function $\tilde U(q,\tilde z)$ for the dimensionless turbulent spectrum $\tilde W(q,\tilde
z)$,
\begin{equation}\label{til_U}
\tilde U=v_{\rm A}kW/U_* \equiv e^{\tilde z-q}\tilde W\,,
\end{equation}
normalized by the energy density flux of CR sources
\begin{equation}
U_*=\frac13S_{*}(m_pc^2)^2\,.
\end{equation}
The expression in parentheses on the l.h.s.\ of Equation~(\ref{N}), to be denoted as $e^{q}\tilde S(q,\tilde z)$, is the
total (diffusion + advection) differential flux of CRs per unit momentum,
\begin{equation}
S=S_{\rm diff}+S_{\rm adv}\,,
\end{equation}
normalized by the source flux $cS_{*}$ and multiplied with $e^{q}$. The dimensionless scale factor for the diffusion flux,
\begin{equation}
\tilde D=\frac{B}{4\pi^2eS_{*}z_n m_pc}\,,
\end{equation}
is a very small number: for parameters chosen in this paper, $\tilde D\sim3\times10^{-6}$.

The boundary conditions at $z=0$ are determined by the source spectrum of CRs and by the turbulent spectrum in the Galactic
disk. In the dimensionless form, we have
\begin{eqnarray}
&&\tilde S(q,0)=e^{-\gamma q}\,, \label{BC_S}\\
&&\tilde U(q,0)=\tilde U_0e^{(\beta-1)q}\,, \label{BC_U}
\end{eqnarray}
where the first condition immediately follows from Equation~(\ref{source}). The second condition is obtained from
Equations~(\ref{D_disk}) and (\ref{ddif}), which gives the disk turbulent spectrum $W_0(q)$. Using Equation~(\ref{til_U}),
we obtain the dimensionless amplitude
\begin{equation}\label{U_0}
\tilde U_0=\frac{Bv_{\rm A}^0}{2\pi^2eS_*D_*m_pc}\equiv \frac{2v_{\rm A}^0z_n}{D_*}\tilde D\,.
\end{equation}
We see that the relative magnitude of the disk turbulence is very small, too. We also note that the ratio $\tilde U_0/\tilde
D$ turns out to be of the order of unity (see Section~\ref{nsol}, Table~1).

\subsection{Approximate model of the CR halo}
\label{approx}

The term on the r.h.s.\ of Equation~(\ref{U}) describes excitation of MHD waves by the outgoing CR flux. Since the CR source
spectrum is a decreasing function of $p$, the excitation is stronger at larger $k$, according to the resonance condition
(\ref{res}). Therefore, the relative magnitude of the excited turbulence as compared to the disk turbulence should {\it
increase} with $k$. Furthermore, adiabatic losses for turbulence, represented by the second term on the l.h.s.\ of
Equation~(\ref{U}), lead to a {\it redshift} as the turbulent spectrum advects upward. Hence, for any given $k$ the excited
turbulence is expected to dominate starting from a certain (momentum-dependent) height, reducing $S_{\rm diff}\propto
W^{-1}$ with respect to $S_{\rm adv}\propto v_{\rm A}$. This makes advection the dominant regime of CR transport at larger
$z$, with the total flux $S\approx S_{\rm adv}= v_{\rm A}N$.

\subsubsection{Advection solution}

One can easily derive the advection solution of Equations~(\ref{N}) and (\ref{U}), by neglecting the diffusion term in the
total flux. Introducing in Equations~(\ref{N}) an auxiliary function for the CR density, $F(q,\tilde z)= e^{\tilde
z+q}\tilde N(q,\tilde z)$, we readily obtain a general solution $F(q+\frac13\tilde z)$. For the flux boundary condition
(\ref{BC_S}), this gives the following CR density:
\begin{equation}\label{N_adv}
\tilde N(q,\tilde z)=e^{-\gamma q-\frac{\gamma+2}3\tilde z}\equiv \tilde N_0^{\rm adv}(q)e^{-\frac{\gamma+2}3\tilde z}\,,
\label{21}
\end{equation}
where $\tilde N_0^{\rm adv}(q)=e^{-\gamma q}$ is the disk spectrum in the advection regime. Substituting this in the r.h.s.\
of Equation~(\ref{U}), we obtain a general solution $\tilde U_0(q-\tilde z)+\int_0^{\tilde z}f(q-\tilde z+x,x)\:dx$, where
$f(q,\tilde z)$ is the resulting r.h.s.\ term. This yields
\begin{eqnarray}
\tilde U(q,\tilde z)=\tilde U_0e^{(\beta-1)(q-\tilde z)}\nonumber\hspace{3.5cm}\\
+\frac{\gamma+2}{4\gamma-7}\left(1-e^{-\frac{4\gamma-7}3\tilde z}\right)
e^{-(\gamma-2)(q-\tilde z)}\nonumber\\
\equiv \tilde U_1(q,\tilde z)+\tilde U_2(q,\tilde z)\,.\label{U_adv}
\label{uzq}
\end{eqnarray}
We see that, due to adiabatic losses the excited turbulence ($\tilde U_2$) increases with height, while the disk turbulence
($\tilde U_1$) decreases. Of course, from Equation~(\ref{til_U}) it follows that the physical spectrum asymptotically
decreases with height as $W(q,\tilde z)\propto e^{-(3-\gamma)\tilde z}$, and thus the diffusion coefficient,
Equation~(\ref{ddif}), correspondingly increases.

A boundary of the advection regime in the $(q,\tilde z)$ plane can be roughly estimated by inserting the derived CR density
and turbulent spectrum into condition $S_{\rm adv}\gtrsim S_{\rm diff}$. Substituting Equation~(\ref{N_adv}) gives
\begin{equation}\label{transition}
e^{-q}\tilde U(q,\tilde z) \gtrsim \frac{\gamma+2}3\tilde D\,.
\end{equation}
For turbulent spectrum (\ref{U_adv}), a simple analysis shows that $U_1$ can be omitted at any $q$ and the advection domain
is determined by $U_2$ only. We obtain $\tilde z\gtrsim \tilde De^{(\gamma-1)q}$ at lower CR energies, where $\tilde
De^{(\gamma-1)q}$ is small; for $\tilde D\sim3\times10^{-6}$, this roughly corresponds to energies $\lesssim10^4$~GeV
($q\lesssim 10$). At higher energies, the advection operates at $(\gamma-2)\tilde z \gtrsim \ln\left(\frac{4\gamma-7}3\tilde
D\right)+(\gamma-1)q$.

\subsubsection{Halo sheath}

The above analysis indicates that the adiabatic losses are essential at large $z$, shaping both the CR and turbulent spectra
described by Equations~(\ref{N_adv}) and (\ref{U_adv}). For CR energies $\lesssim10^4$~GeV, we found that a transition to
this advection regime occurs in a relatively narrow layer $\tilde z\lesssim1$ -- below we refer to it as the {\it CR halo
sheath}. Hence, we can approximately describe the sheath structure by assuming that the adiabatic losses do not noticeably
affect the total flux within this narrow layer. Then Equation~(\ref{N}) with boundary condition (\ref{BC_S}) is reduced to
\begin{equation}\label{flux}
-\tilde D\frac{e^{q}}{\tilde U}\frac{\partial \tilde N}{\partial \tilde z}+\tilde N\approx e^{-\gamma q-\tilde z}\,,
\label{nvel}
\end{equation}
i.e., the total CR flux in the sheath is equal to the source flux.

The further analysis can be performed for an arbitrary turbulent spectrum $\tilde U(q,\tilde z)$. Solving
Equation~(\ref{flux}) gives the CR density in the sheath,
\begin{eqnarray}
\tilde N(q,\tilde z)= e^{-\gamma q-\tilde z}-\delta \tilde N_0(q)e^{\tilde V(q,\tilde z)}\nonumber\hspace{2.cm}\\
+e^{-\gamma q+\tilde V(q,\tilde z)}\int_0^{\tilde z}e^{-\tilde V(q,x)-x}\:dx\,,\label{sheath}
\label{13}
\end{eqnarray}
where
\begin{equation}\label{V}
\tilde V(q,\tilde z)=\frac{e^{-q}}{\tilde D}\int_0^{\tilde z}\tilde U(q,x)\:dx\,.
\end{equation}
We see that the CR spectrum in the Galactic disk ($z=0$),
\begin{equation}
\tilde N(q,0)= e^{-\gamma q}-\delta \tilde N_0(q)\equiv \tilde N_0(q)\,,
\end{equation}
is characterized by unknown function $\delta \tilde N_0(q)$, describing deviation from the advection spectrum. The negative
sign of this term follows from Equation~(\ref{flux}), as the relative contribution of diffusion to the total flux increases
with energy. The value of $\delta \tilde N_0(q)$ is obtained by matching the sheath spectrum (\ref{13}) with the advection
spectrum (\ref{N_adv}) at the unknown sheath edge $\tilde z_{\rm sh}(q)$. Continuity of $\tilde N$ and $\partial \tilde
N/\partial \tilde z$ yields equation for $\tilde z_{\rm sh}(q)$,
\begin{equation}\label{z_sh}
e^{-q}\tilde U(q,\tilde z_{\rm sh})\left(e^{\frac{\gamma-1}3\tilde z_{\rm sh}}-1\right)= \frac{\gamma+2}3\tilde D\,,
\end{equation}
which determines the density deviation,
\begin{eqnarray}
\frac{\delta \tilde N_0(q)}{e^{-\gamma q}}=\left(1-e^{-\frac{\gamma-1}3\tilde z_{\rm sh}}\right)
e^{-\tilde V(q,\tilde z_{\rm sh})-\tilde z_{\rm sh}}\nonumber\hspace{1cm}\\
+\int_0^{\tilde z_{\rm sh}}e^{-\tilde V(q,x)-x}\:dx\,.\label{delta_N}
\end{eqnarray}

Properties of the halo sheath can be understood by employing Equation~(\ref{U_adv}), which accurately describes the
turbulent spectrum both in the Galactic disk and in the advection regime. The resulting approximate solution for the sheath
edge,
\begin{equation}\label{z_sh2}
\tilde z_{\rm sh}(q) \approx \left\{\begin{array}{ll}\sqrt{\frac3{\gamma-1}\tilde D e^{(\gamma-1)q}}\,, & \tilde z_{\rm sh}\lesssim1;
\\[.3cm]
\frac3{4\gamma-7}\left[\ln\left(\frac{4\gamma-7}3\tilde D\right)+(\gamma-1)q\right],  & \tilde z_{\rm sh}\gtrsim1,\end{array}\right.
\end{equation}
gives the expressions valid for small and large values of $\tilde z_{\rm sh}$, respectively. Equation~(\ref{z_sh2})
shows that the disk turbulence is unimportant for the transition to advection regime. We also note that the energy at which
$\tilde z_{\rm sh}(q)\sim1$ is about that deduced from Equation~(\ref{transition}). Substituting Equation~(\ref{z_sh2}) for
lower energies, where $\tilde z_{\rm sh}(q)\ll1$, into Equation~(\ref{V}) gives $\tilde V(q,\tilde z_{\rm sh})\sim1$.
Therefore, from Equation~(\ref{delta_N}) we conclude that the relative deviation from the advection spectrum is small at
lower energies, $\delta \tilde N_0/e^{-\gamma q}\sim \tilde z_{\rm sh}$. On the other hand, for sufficiently high energies
the excitation becomes inefficient and the disk turbulence $\tilde U_1(q,\tilde z)$ in Equation~(\ref{U_adv}) dominates up
to $\tilde z\gtrsim1$. This occurs for $q\gtrsim q_U$, as determined from
\begin{equation}\label{q_U}
q_U=\frac{\ln\left(\frac{\gamma+2}{4\gamma-7}\tilde U_0^{-1}\right)}{\gamma+\beta-3}\,.
\end{equation}
Equation~(\ref{delta_N}) can be rewritten in a different form, more convenient for the analysis at high energies.
Integrating the second term by parts yields the CR spectrum in the Galactic disk,
\begin{eqnarray}
\frac{\tilde N_0(q)}{e^{-\gamma q}}=e^{-\tilde V(q,\tilde z_{\rm sh})-\frac{\gamma+2}3\tilde z_{\rm sh}}\nonumber\hspace{3.cm}\\
+\frac{e^{-q}}{\tilde D}\int_0^{\tilde z_{\rm sh}}\tilde U(q,x)e^{-\tilde V(q,x)-x}\:dx\,.\label{N_disk}
\end{eqnarray}
For high energies, where $\tilde z_{\rm sh}(q)\gg1$, Equation~(\ref{V}) gives $\tilde V(q,\tilde z_{\rm sh})\ll1$. Then,
substituting $\tilde U(q,\tilde z)\approx \tilde U_1(q,\tilde z)$ in Equation~(\ref{N_disk}) we derive the CR spectrum for
diffusion regime,
\begin{equation}\label{N_diff}
\tilde N_0^{\rm diff}(q)=\frac{\tilde U_0}{\beta\tilde D}e^{-(\gamma+2-\beta)q}\,.
\end{equation}
This diffusion asymptote can also be straightforwardly derived by dropping the advection term $\tilde N$ in
Equation~(\ref{flux}) and integrating it with $\tilde U(q,\tilde z)=\tilde U_1(q,\tilde z)$.

\subsection{CR spectrum in the Galactic disk} \label{GD_spectrum}

\begin{figure}
\begin{center}
\resizebox{\hsize}{!}{\includegraphics{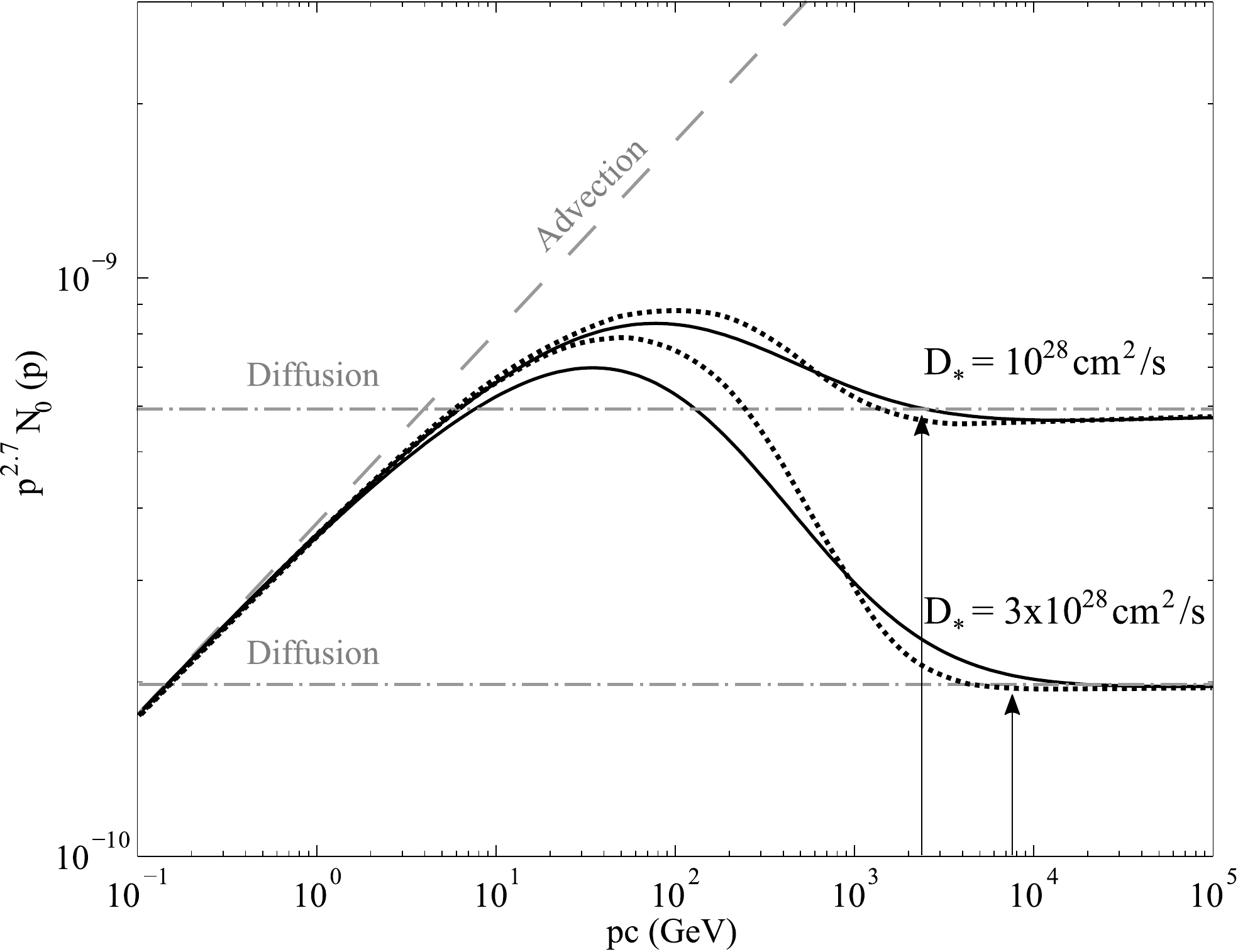}}
\caption{CR spectrum in the Galactic disk. The dotted lines represent approximate analytical solution (\ref{N_approx})
with $q=\ln(p/m_pc)$, obtained for $\gamma=2.3$, $\beta=1.6$, and the two values of $D_*$. Advection asymptote (\ref{N_adv})
and diffusion asymptote (\ref{N_diff}) for low and high energies, respectively, are also plotted. The vertical arrows
indicate the positions where $q=q_U$, Equation~(\ref{q_U}). The solid lines show the exact result of numerical solution of
Equations~(\ref{N}) and (\ref{U}).} \label{fig2}
\end{center}
\end{figure}

The approximate halo model provides an analytical expression for the CR spectrum in the disk, Equations~(\ref{N_disk}). One
can further simplify this expression by assuming that the CR diffusion in the halo sheath is primarily controlled by the
disk turbulence, $\tilde U\approx \tilde U_1$. In order to obtain a convenient tractable formula, we approximate $\tilde
U_1(q,\tilde z)$ in Equation~(\ref{U_adv}) by $U_0e^{(\beta-1)q}\theta(\tilde z_\infty-\tilde z)$, where $\theta(x)$ is the
Heaviside step function and $\tilde z_\infty$ is to be determined. By virtue of Equation~(\ref{V}) we get $\tilde V(q,\tilde
z)=\kappa\tilde z$ for $\tilde z\leq \tilde z_\infty$ and $\tilde V(q,\tilde z)=\kappa \tilde z_\infty$ otherwise, where
\begin{equation}
\kappa(q)=\frac{\tilde U_0}{\tilde D}e^{-(2-\beta)q}\,,
\end{equation}
does not exceed unity for relativistic protons, as follows from Equation~(\ref{U_0}). Substituting this in
Equation~(\ref{N_disk}) gives
\begin{equation}\label{N_approx}
\frac{\tilde N_0(q)}{e^{-\gamma q}}=e^{-\kappa \tilde z_{\rm sh}^{\rm m}-\frac{\gamma+2}3\tilde z_{\rm sh}}
+\frac{\kappa}{\kappa+1}\left(1-e^{-(\kappa+1)\tilde z_{\rm sh}^{\rm m}}\right),
\label{n0q}
\end{equation}
with $\tilde z_{\rm sh}^{\rm m}=\min(\tilde z_{\rm sh},\tilde z_\infty)$. We determine $\tilde z_\infty$ from the condition
that at high energies Equation~(\ref{N_approx}) should converge to diffusion asymptote (\ref{N_diff}). Taking into account
that $\tilde z_{\rm sh}(q)\gg1$ and $\kappa(q)\ll1$ in this case, we obtain $\tilde z_\infty=-\ln(1-1/\beta)$.

Figure~\ref{fig2} compares Equation~(\ref{N_approx}) with the disk spectrum obtained from numerical solution of
Equations~(\ref{N}) and (\ref{U}) (in Section~\ref{nsol} we describe the approach used to solve the governing equations in
general case). The approximate analytical solution shows a remarkable agreement across the whole energy range between the
advection and diffusion asymptotes. We also see that the transition to diffusion asymptote (\ref{N_diff}) occurs near
$q\approx q_U$, as predicted by Equation~(\ref{q_U}).

Thus, Figure~\ref{fig2} demonstrates that the CR spectrum in the disk forms as a result of nonlinear interplay between
advection and diffusion. The former regime, controlled by the excited turbulence, completely dominates at energies below a
few GeV, where CR spectrum (\ref{N_adv}) follows that of the source. On the other hand, at sufficiently high energies above
$\sim m_pc^2\exp(q_U)$, CR spectrum (\ref{N_diff}) forms as a result of diffusion prescribed by the disk turbulence, because
it takes longer for the excited turbulence to come up. Thus, energetic CRs become trapped in an extended sheath, causing the
formation of an ``excess'' seen at intermediate energies. The excess provides a smooth transition between the advection and
diffusion asymptotes.

The sheath edge $z_{\rm sh}(q)$, given by Equation~(\ref{z_sh2}), can be considered as the ultimate size of the CR halo,
beyond which CR transport becomes advective at any energy. On the other hand, we also note that CR spectrum (\ref{N_diff}),
representing diffusive regime at high energies, coincides with the spectrum predicted by the model of a static halo with the
physical size of $2z_n/\beta$. Thus, the the physical size of the CR halo, $z_{\rm H}$, can be defined as
\begin{equation}\label{z_H}
z_{\rm H}(q)=\min\left\{z_{\rm sh}(q),\frac{2z_n}{\beta}\right\}.
\end{equation}
where $z_{\rm sh}(q)$ is approximated by the first line of Equation~(\ref{z_sh2}).

\section{Numerical solution}\label{nsol}

In this section we numerically solve a system of non-stationary governing equations (\ref{tdif}) and (\ref{vaw}) for CR
spectrum $N(p,z,t)$ and spectrum of MHD waves $W(k,z,t)$, until a steady-state solution is reached. By replacing the
boundary condition for the total differential flux of CRs at $z=0$ with the source delta-function, we obtain the following
equations:
\begin{eqnarray}
\frac{\partial N}{\partial t}+\frac{\partial}{\partial z}\left(v_{\rm adv}N-D\frac{d N}{\partial z}\right)
-\frac13\frac{d v_{\rm adv}}{dz}\,\frac{\partial}{\partial p}\left(pN\right)\nonumber\\
=2S_*\left(\frac{m_pc}{p}\right)^{\gamma}\delta(z)\,,
\label{tdif1}\\
\frac{\partial W}{\partial t}+\frac{\partial}{\partial z}(v_{\rm A}W)-\frac{d v_{\rm A}}{d z}\,\frac{\partial}{\partial k}(kW)
=2\Gamma W\,,\hspace{.85cm}\label{vaw1}
\end{eqnarray}
where $p$ and $k$ are related by resonance condition (\ref{resonant}), $D(p,z)$ and $\Gamma(k,z)$ are given by
Equations~(\ref{ddif}) and (\ref{gmm}), respectively, and $v_{\rm A}(z)<c$ is assumed. While the disk turbulence is
isotropic, the excited turbulence propagates only upward. Therefore, the advection velocity $v_{\rm adv}(z)$ should
gradually approach the local Alfv\'{e}n velocity, which is approximated by
\begin{equation}\label{vc}
v_{\rm adv}(z) = v_{\rm A}(z)\tanh(z/\Delta z)\,.
\end{equation}
Here $\Delta z$ is the characteristic height of the source distribution in the disk (of the order of the disk
half-thickness). The problem is solved for two characteristic models of the gas density profile, described by
Equations~(\ref{n12}) and (\ref{beta}) and illustrated in Figure~\ref{fig1}.

A steady-state solution is obtained for initial conditions $N(p,z,0)=0$ and $W(k,z,0)=W_0(k)$. Boundary conditions for the
CR spectrum are
\begin{eqnarray}
&&N(p_{\rm max}, z,t) = 0 \,, \\
&&\left.\left(v_{\rm adv}N-D\frac{\partial N}{\partial z}\right)\right|_{z = z_{\rm max}}=cN \,,
\end{eqnarray}
where we set $p_{\rm max}=10^7m_pc$. The second condition defines the position $z = z_{\rm max}$ where CRs escape from the
halo with luminal velocity. Boundary conditions for the MHD spectrum are
\begin{eqnarray}
&&W(k_{\rm max}, z,t) = 0 \,, \\
&&W(k, 0,t) = W_0(k) \,,
\end{eqnarray}
where $k_{\rm max}$ corresponds to the resonant momentum of $0.1m_pc$.

The following input parameters for the model are estimated from available observations: the differential flux of CR sources
in the disk $S(E)$, the disk spectrum of MHD turbulence $W_0(k)$, the density profile of ionized gas $n(z)$, and the disk
half-thickness $\Delta z$. For the disk turbulence, we assume a Kolmogorov spectrum with $\beta=1.67$. The magnetic field
strength is set to $B=2~\mu$G, so that the Alfv\'{e}n velocity in the disk $v_{\rm A}^0$ (determined by the gas density at
$z=0$) is fixed.

As discussed in Section~\ref{GD_spectrum}, different parts of the CR spectrum depend on different parameters. Below several
GeV the energy spectrum in the disk can be approximated by $N_0^{\rm adv}(E) = S(E)/v_{\rm A}^0$; at energies above several
TeV the spectrum approaches $N_0^{\rm diff}(E) = (2z_n/\beta) S(E)/D_0(E)$. Thus, one can evaluate $S_*$ and $\gamma$ (for a
given gas density profile) by fitting the CR spectrum at low energies, and fitting at high energies yields $D_*$.

\begin{figure}
\begin{center}
\resizebox{\hsize}{!}{\includegraphics{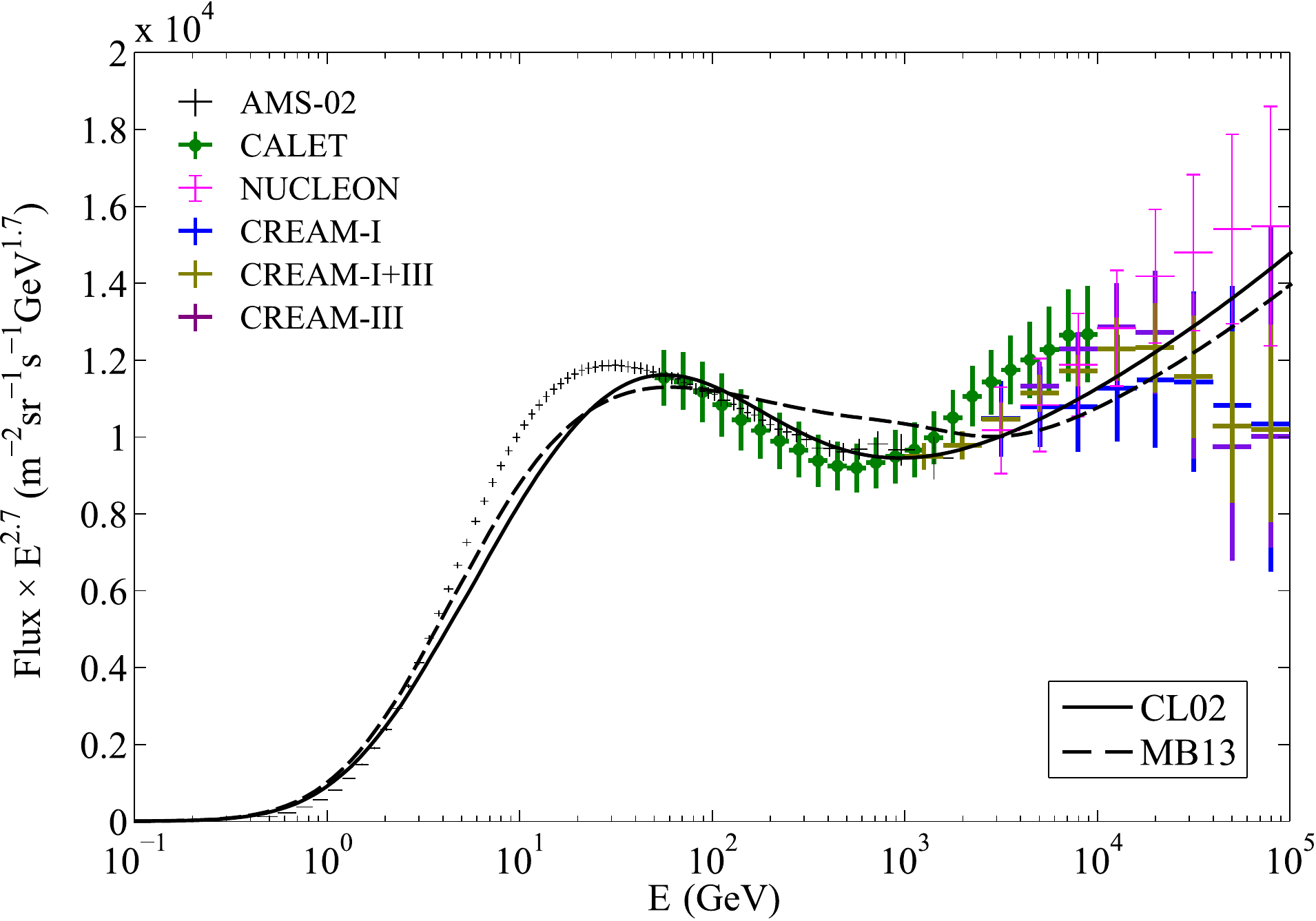}}
\caption{Comparison of the numerically calculated proton spectrum in the Galactic disk with observation data. The
differential flux $vN_0(E)/(4\pi)$ is multiplied with $E^{2.7}$. The numerical results are for the gas density models given
by Equation~(\ref{n12}) (CL02) and Equation~(\ref{beta}) (MB13), the data are taken from different observations indicated
in the legend.} \label{fig3}
\end{center}
\end{figure}

The CR spectrum at intermediate energies depends both on $S_*$ and $D_*$. Here we need to reproduce both the spectral break
at about 0.6~TeV and the spectral shape between $\sim10$~GeV and $\approx0.6$~TeV. It turns out that the observed spectrum
in this energy range favors values of $S_*$ higher than those deduced from low-energy fitting. We resolve this issue by
introducing an additional parameter, the disk half-thickness $\Delta z=100$~pc. Equation~(\ref{vc}) shows that the CR
advection decreases with $\Delta z$, and thus $N_0(E)$ becomes larger at low energies. This allows us to increase $S_*$
without changing the value of $v_{\rm A}^0$.

\begin{table}[!h]
\begin{center}
\caption{Model parameters obtained from data fitting.\\ Fixed parameters are $\beta=1.67$ and $B=2~\mu$G.}
\begin{tabular}{cccccc} \tableline\tableline
Gas            & $v_{\rm A}^0$   & $S_*$                           & $\gamma$ & $D_*$              \\
density        & (cm~s$^{-1}$)   & (cm$^{-2}$~s$^{-1}$~GeV$^{-1}$) &          & (cm$^2$~s$^{-1})$  \\
\tableline
CL02$^{\rm a}$ & $1.2\times10^6$ & $5.5\times10^{-4}$              & 2.25     & $5.3\times10^{27}$ \\
MB13$^{\rm b}$ & $6.0\times10^5$ & $4.4\times10^{-4}$              & 2.25     & $5.3\times10^{27}$ \\
\tableline
\end{tabular}
\tablenotetext{a}{Equation~(\ref{n12}), from \citet{cord03}.} \tablenotetext{b}{Equation~(\ref{beta}), from
\citet{mill16}.}\tablenotetext{}{}
\end{center}
\end{table}

The proton spectrum in the disk is plotted in Figure~\ref{fig3}, showing our numerical results and data from \citet{agu15}
(AMS-02), \citet{calet19} (CALET), \citet{nucleon19} (NUCLEON), \citet{cream11} (CREAM-I), and \citet{cream17}
(CREAM-I+III). The data were collected using Cosmic-Ray DataBase (CRDB v4.0) by \citet{Maurin2020}. The parameters
needed to reproduce the data are presented in Table~1. In order to compare our results with the observations, the solar
modulation of numerically calculated spectrum was artificially introduced by using the force-field approximation with
potential $\phi = 0.5$~GV \citep{gleeson}.

\section{Discussion and conclusions}
\label{discuss}

Figure~\ref{fig3} shows that at energies $\gtrsim50$~GeV the exponential profile (CL02) of ionized gas density provides a
remarkably good fit for the Galactic spectrum of protons, reproducing the spectral shape and the position of spectral break.
The curve for the power-law profile (MB13) noticeably deviates from the data, but still shows a qualitative agreement. This
suggests that the spectral shape is substantially affected by the spatial distribution of gas. We point out that both
exponential and power-law profiles are just model approximations, and certain features of the proton spectrum may therefore
be introduced by (unknown) details of the actual gas distribution.

Both theoretical curves substantially deviate from experimental data at low energies, even though the chosen value of the
disk half-thickness is already relatively high, $\Delta z=0.1$~kpc. We were unable to reproduce the low-energy spectrum even
by pushing it to $\Delta z=0.3$~kpc. Several factors can explain this deviation:
\begin{enumerate}
\item The solar modulation modifies the low-energy spectrum, and the used force-field approximation could be too crude
    for our purposes. As one can see from Table~1, the relative variation of $S_*$ between models CL02 and MB13 is much
    smaller than that of $v_{\rm A}^0$.
\item The solution favors smaller $v_{\rm A}^0$, and the used values may be higher than the actual values in the
    Galactic disk.
\item We consider a simple one-dimensional problem. However, realistic models should include horizontal inhomogeneities
    -- in particular those due to coexistence of phases of the warm ionized and warm neutral media in the Galactic
    disk. The latter leads to significant inhomogeneities in $v_{\rm A}^0$ and, therefore, is expected to affect the
    advection-dominated transport of low-energy protons. Furthermore, a three-dimensional structure of the magnetic
    field near the disk may have an impact on the quantitative results.
\item We do not consider stochastic re-acceleration which could be important for low-energy protons, increasing the
    magnitude of the spectrum.
\end{enumerate}
Point 3 represents the main conceptual simplification adopted in our model. Inclusion of disk inhomogeneities is a highly
nontrivial problem on its own right, which requires a separate careful analysis. On the other hand, point 4 can be, in
principle, incorporated; in the present work CR re-acceleration is omitted as we focus on the essential processes governing
the halo formation.

One of the main conclusions of our paper is that the CR halo can be naturally formed without a turbulent cascade, which is
in contrast to recent studies by \citet{evoli18}. Both models include CR-generated MHD waves and their advection from the
Galactic disc as critically important ingredients. However, Evoli's model also includes cascading of advected turbulence to
larger $k$, which causes CR scattering to reduce with height and thus sets the halo's edge beyond which CRs can freely
escape. On the contrary, in Appendix~\ref{notes} we show that the cascading process in the halo could be practically
disabled for relevant $k$. Instead, we point out the importance of adiabatic losses in wave equation (\ref{vaw}), leading to
the turbulent redshift and thus {\it enhancing} CR scattering by the self-generated MHD waves.

We demonstrate that the size $z_{\rm H}(E)$ of the CR halo at lower energies is determined by the {\it halo sheath}, an
energy-dependent region around the Galactic disk beyond which a transition to Alfv\'{e}nic advection occurs. At sufficiently
high energies $z_{\rm H}$ is set by the characteristic thickness of the ionized gas distribution, i.e., becomes
energy-independent. The resulting value of $z_{\rm H}(E)$ is approximated by Equation~(\ref{z_H}). Thus, unlike other
models, we conclude that the CR halo is about 1~kpc or less, depending on the energy. This favors local origins of CRs in
the disk \citep[see][]{breit02}, suggesting that a global CR mixture in the Galaxy is insignificant.

Similar to the conclusion drawn by \citet{Blasi2012} and \citet{evoli18}, we show that the CR spectrum in a broad energy
range between several GeV and TeV is shaped by a competition between advection and diffusion. The advection regime with
spectrum $N_0(E)\propto E^{-\gamma}$ at lower energies is controlled by the self-excited turbulence, while at high energies
CR spectrum $N_0(E)\propto E^{-(\gamma+2-\beta)}$ forms as a result of diffusion prescribed by the disk turbulence. The
width of the energy range where a transition between these two regimes occurs is proportional to $\propto (S_*D_*/Bv_{\rm
A}^0)^{\frac1{\gamma+\beta-3}}$, as follows from Equations~(\ref{U_0}) and (\ref{q_U}) and illustrated in Figure~\ref{fig2}.
This transition naturally explains the observed spectral break.

In a forthcoming publication we plan to apply the presented model to other CR species, including primary and secondary
nuclei. This will allow us to compare theoretical spectra of important isotopes with observations and, thus, verify the
viability of the model.

\section*{Acknowledgments}

The authors are grateful to Pasquale Blasi, Carmelo Evoli, Sarah Recchia, and Andy Strong for useful discussions. The work
is supported by Russian Science Foundation via the Project 20-12-00047.

\bibliographystyle{apj}
\bibliography{halo_text}

\appendix
\section{Appendix A: Notes on wave equation~(10)}
\label{notes}

{\it ``Adiabatic losses''.} The origin of the second term on the l.h.s.\ of Equation~(\ref{vaw}) is the third term in
Equation~(\ref{Liu1}). It represents the process which is analogous to the adiabatic losses for CRs, described by the last
term in Equation~(\ref{tdif}). As a result, the advected turbulent spectrum shifts toward smaller $k$ due to increase of the
Alfv\'{e}n velocity with height. The factor of 1/3, emerging in Equation~(\ref{tdif}) due to isotropy of the CR spectrum, is
missing in Equation~(\ref{vaw}) because waves are assumed to propagate along the $z$-axis. In Section~\ref{mech} we show
that this new term in the wave equation is crucial for the halo formation.

{\it Regular damping.} Following \citet{brag65} and adopting his notations, viscous damping of non-compressive MHD waves is
determined by the viscosity perpendicular to the magnetic field, $\eta_\perp\sim \eta_0/(\Omega\tau)^2$, where $\eta_0\sim
10^{-12}nT\tau$ is the gas-kinetic viscosity of thermal ions (in g~cm$^{-1}$ s$^{-1}$ and $T$ in eV), $\Omega\sim10^4\,B$ is
their gyrofrequency (in s$^{-1}$), and $\tau\sim 10^6\,T^{3/2}/n$ is the ion-ion collision time (in s). This implies a
reduction of the gas-kinetic viscosity by a factor of $(\Omega\tau)^2$, estimated to be of the order of $\sim10^{16}$ or
larger for conditions assumed outside the disk. By comparing the resulting damping rate, $\sim (\eta_{\perp}/mn)k^2$, versus
the adiabatic rate, $\sim v_{\rm A}/z_n$, we conclude that the former is completely negligible for any relevant $k$.
Similarly, we obtain that the resistive damping is completely negligible, too.

{\it Nonlinear terms.} A cascade or/and nonlinear damping of non-compressive MHD waves may occur due to coupling between
waves $W_+$ and $W_-$ traveling in opposite directions with respect to the magnetic field \citep{skil2, verma}. The
functional form of the resulting nonlinear term(s) -- being a subject of long ongoing debates -- strongly depends on
physical conditions and the dominant coupling mechanism \citep[see, e.g.,][and references
therein]{Goldreich1995,Goldreich1997,verma,ptus03,ptus06}.

For Alfv\'{e}n waves propagating upward ($W_+$) and downward ($W_-$), their coupling could be mediated by sound waves
\citep{skil2}. However, sound waves are relatively heavily damped due to gas-kinetic viscosity $\eta_0$ \citep{brag65}, and
therefore only propagate if their frequency $c_{\rm s}k$ exceeds the damping rate $\sim(\eta_0/mn)k^2$. Hence, such a
cascade cannot operate at large $k$ due to the lack of sound waves.

For our conditions, sound waves are able to propagate at relatively small $k$, corresponding to the resonant CR energies of
$\gtrsim10^{14}$~eV. The wave coupling results in a cascade term $\propto W_+\partial(kW_-)/\partial k$ on the r.h.s.\ of
Equation~(\ref{vaw}) written for $W_+$ \citep[and vice versa for $W_-$, see][]{skil2}. Below we show (see
Section~\ref{approx}) that the energy density of waves at such small $k$ is determined by the disk turbulent spectrum, i.e.,
$W_+(k)\propto k^{-\beta}$ with $\beta\approx5/3$. Hence, the cascade term for waves $W_-$, proportional to
$\propto\partial(kW_+)/\partial k$, is negative, which results in their damping. Taking into account additional damping of
$W_-$ waves by CRs and assuming no wave sources in the halo, we can safely set $W_-=0$. This allows us to omit the cascade
term also for $W_+$ waves.

As regards the cascade of purely incompressible Alfv\'{e}nic turbulence \citep[][]{Goldreich1995,verma,ptus03}, it cannot
create waves with wavevectors (along the magnetic field) which were not preset initially \citep[][]{Goldreich1997}. Assuming
no sources of $W_-$ waves in the halo, we therefore neglect the cascade term for $W_+$ waves, too.


\end{document}